\documentclass[prl,twocolumn,superscriptaddress]{revtex4}
\usepackage{graphicx}
\usepackage{dcolumn}

\input{psfig}

\def\pmz{\pi^+\pi^-\pi^0}
\begin{document}

\onecolumngrid
\begin{center}
\vskip -1.0cm 
{\large\bf A Measurement of the $K_L$ Charge Asymmetry } \\
\vskip .2cm
(Submitted to Physical Review Letters, December 31st, 2001)
% Version 3.0 --- version of submission to prl
\end{center}
\lineskip 1.5pt
A.~Alavi-Harati$^{12}$,
T.~Alexopoulos$^{12}$,
M.~Arenton$^{11}$,
K.~Arisaka$^2$,
S.~Averitte$^{10}$,
R.F.~Barbosa$^{7,**}$,
A.R.~Barker$^5$,
M.~Barrio$^4$,
L.~Bellantoni$^7$,
A.~Bellavance$^9$,
J.~Belz$^{10}$,
R.~Ben-David$^{7}$,
D.R.~Bergman$^{10}$,
E.~Blucher$^4$, 
G.J.~Bock$^7$,
C.~Bown$^4$, 
S.~Bright$^4$,
E.~Cheu$^1$,
S.~Childress$^7$,
R.~Coleman$^7$,
M.D.~Corcoran$^9$,
G.~Corti$^{11}$, 
B.~Cox$^{11}$,
M.B.~Crisler$^7$,
A.R.~Erwin$^{12}$,
R.~Ford$^7$,
A.~Glazov$^4$,
A.~Golossanov$^{11}$,
G.~Graham$^{4}$, 
J.~Graham$^4$,
K.~Hagan$^{11}$,
E.~Halkiadakis$^{10}$,
J.~Hamm$^1$,
K.~Hanagaki$^{8}$,  
S.~Hidaka$^8$,
Y.B.~Hsiung$^7$,
V.~Jejer$^{11}$,
D.A.~Jensen$^7$,
R.~Kessler$^4$,
H.G.E.~Kobrak$^{3}$,
J.~LaDue$^5$,
A.~Lath$^{10}$,
A.~Ledovskoy$^{11}$,
P.L.~McBride$^7$,
P.~Mikelsons$^5$,
E.~Monnier$^{4,*}$,
T.~Nakaya$^{7}$,
K.S.~Nelson$^{11}$,
H.~Nguyen$^{7,\dagger}$,
V.~O'Dell$^7$, 
M.~Pang$^7$, 
R.~Pordes$^7$,
V.~Prasad$^4$,
X.R.~Qi$^7$, 
B.~Quinn$^{4,\S}$,
E.J.~Ramberg$^7$, 
R.E.~Ray$^7$,
A.~Roodman$^{4}$, 
M.~Sadamoto$^8$, 
S.~Schnetzer$^{10}$,
K.~Senyo$^{8}$, 
P.~Shanahan$^7$,
P.S.~Shawhan$^{4}$,
J.~Shields$^{11}$,
W.~Slater$^2$,
N.~Solomey$^4$,
S.V.~Somalwar$^{10}$, 
R.L.~Stone$^{10}$, 
E.C.~Swallow$^{4,6}$,
S.A.~Taegar$^1$,
R.J.~Tesarek$^{10}$, 
G.B.~Thomson$^{10}$,
P.A.~Toale$^5$,
A.~Tripathi$^2$,
R.~Tschirhart$^7$,
S.E.~Turner$^2$, 
Y.W.~Wah$^4$,
J.~Wang$^1$,
H.B.~White$^7$, 
J.~Whitmore$^7$,
B.~Winstein$^4$, 
R.~Winston$^4$, 
T.~Yamanaka$^8$,
E.D.~Zimmerman$^{4}$
% \lineskip 1pt
% \vspace*{.1 in} 
\footnotesize
\begin{center}
\centerline{ \large The KTeV Collaboration}
\vspace*{.1 in} 
\it
$^1$ University of Arizona, Tucson, Arizona 85721 \\
$^2$ University of California at Los Angeles, Los Angeles, California 90095 \\
$^{3}$ University of California at San Diego, La Jolla, California 92093 \\
$^4$ The Enrico Fermi Institute, The University of Chicago, 
Chicago, Illinois 60637 \\
$^5$ University of Colorado, Boulder, Colorado 80309 \\
$^6$ Elmhurst College, Elmhurst, Illinois 60126 \\
$^7$ Fermi National Accelerator Laboratory, Batavia, Illinois 60510 \\
$^8$ Osaka University, Toyonaka, Osaka 560-0043 Japan \\
$^9$ Rice University, Houston, Texas 77005 \\
$^{10}$ Rutgers University, Piscataway, New Jersey 08854 \\
$^{11}$ The Department of Physics and Institute of Nuclear and 
Particle Physics, \\
University of Virginia, Charlottesville, Virginia 22901 \\
$^{12}$ University of Wisconsin, Madison, Wisconsin 53706 \\
\end{center}
\vspace*{.1 in}

\normalsize

%\begin{abstract}
\vspace{-0.1in}
We present a 
measurement of the charge asymmetry $\delta_L$ in the mode 
$K_L \to \pi^{\pm}e^{\mp}\nu$ based on 298 million analyzed decays.
We measure a value of 
$\delta_L~=~(3322~\pm~58(stat)~\pm~47(sys))\cdot 10^{-6}$, 
in good agreement with previous measurements and 2.4 times more precise than the
current best published result.  
The result is used to place more stringent limits on CPT and 
$\Delta S = \Delta Q$ violation in the neutral kaon system.
\\
\\
\hspace{-0.18in} PACS numbers: 11.30.Er, 11.30.-j, 13.25.Es, 13.20.-v

%\end{abstract}

%\maketitle{}

%\section{Introduction}
\vspace{1.0cm}
\twocolumngrid
The charge asymmetry in $K_L$ semileptonic decays is deeply 
related to CP violation through neutral kaon mixing.
The $K_L$ wavefunction is proportional to 
$(1+\epsilon_L)K^0~-~(1-\epsilon_L)\overline{K^0}$, where $\epsilon_L$ 
parameterizes CP violation in mixing.  This parameter
can be determined in the decay modes $K_L \to \pi^{\pm} e^{\mp} \nu$ 
(Ke3) and $K_L \to \pi^{\pm} \mu^{\mp} \nu$ (K$\mu$3). 
% Assuming that the rule $\Delta{S}=\Delta{Q}$ is obeyed, the states
% $e^+\pi^-\nu$ and $e^-\pi^+\bar{\nu}$ tags the $K^0$ and 
% $\overline{K^0}$ components respectively.
The charge asymmetry ($\delta_L$) is defined as
$$\delta_L = \frac{BR(e^+\pi^-) - BR(e^-\pi^+)}
{BR(e^+\pi^-) + BR(e^-\pi^+)}.$$
Assuming 
$\Delta{S}=\Delta{Q}$ and no CPT violation in 
the decay $K^0(\overline{K^0}) \to e^+\pi^-\nu (e^-\pi^+\bar{\nu})$,
$\delta_L$ is simply 2Re~$\epsilon_L$.
To first order in the parameters that violate CP and CPT,
$$\delta_L = 2Re~\epsilon_L - 2Re~Y - Re(x - \bar{x}).$$
The terms $Y$ and $Re(x-\bar{x})$ parameterize CPT violation in the 
$\Delta{S}=\Delta{Q}$ and $\Delta{S}=-\Delta{Q}$ transitions 
respectively\cite{bd,daphnehb}:
$$\frac{<e^+\pi^-\nu | K^0>}{<e^-\pi^+\bar{\nu} | \overline{K^0}>^*} = 
\frac{1-Y}{1+Y}$$
$$ x = \frac{<e^+\pi^- \nu | \overline{K^0}>}{ <e^+\pi^-\nu | K^0>} ~~~~~~
\bar{x} = \frac{<e^-\pi^+ \bar{\nu} | K^0>^*}{ <e^-\pi^+\bar{\nu} |
\overline{K^0}>^*}.$$
In the Standard Model, the $\Delta{S}=\Delta{Q}$ violation is 
CPT-conserving and occurs in 2nd order weak decays.  
Estimates for $|x|$ are in the
range $10^{-7}$\cite{Dib}. $\delta_L$ is modified to 
order $|x|^2$ and will not be considered in this letter.

The measurement of $\delta_L$ can be compared to expectations derived
from the $K_L \to 2\pi$ (K$\pi$2) 
amplitudes ($\eta_{+-}, \eta_{00}$)\cite{daphnehb,barmin}:
\begin{equation}
\label{cpt}
Re~(Y + \frac{x - \bar{x}}{2} + a) = 
Re(\frac{2}{3}\eta_{+-} + \frac{1}{3}\eta_{00}) - \frac{\delta_L}{2} .
\end{equation}
This comparison is sensitive to 
CPT violation in Ke3 and K$\pi$2 decays, where the latter is 
parameterized by $Re~a$.

The current world average on
$\delta_L$ comes mainly from a measurement by the
CERN-Heidelberg collaboration in 1974.  
Using 34 million Ke3 decays, they found
$\delta_L~=~(3409~\pm~171(stat)~\pm~50(sys))\cdot 10^{-6}$\cite{Geweniger}.
We describe in this letter a measurement of $\delta_L$ in the Ke3 mode
based on approximately 298 million Ke3 decays.  

%\section{Beam and Detector}

\begin{figure}
% \rule{2cm}{0.2mm}\hfill\rule{2cm}{0.2mm}
% \vskip 4cm
% \rule{5cm}{0.2mm}\hfill\rule{2cm}{0.2mm}
% \psfig{figure=detector.eps,height=2.0in,width=3.5in}
\psfig{figure=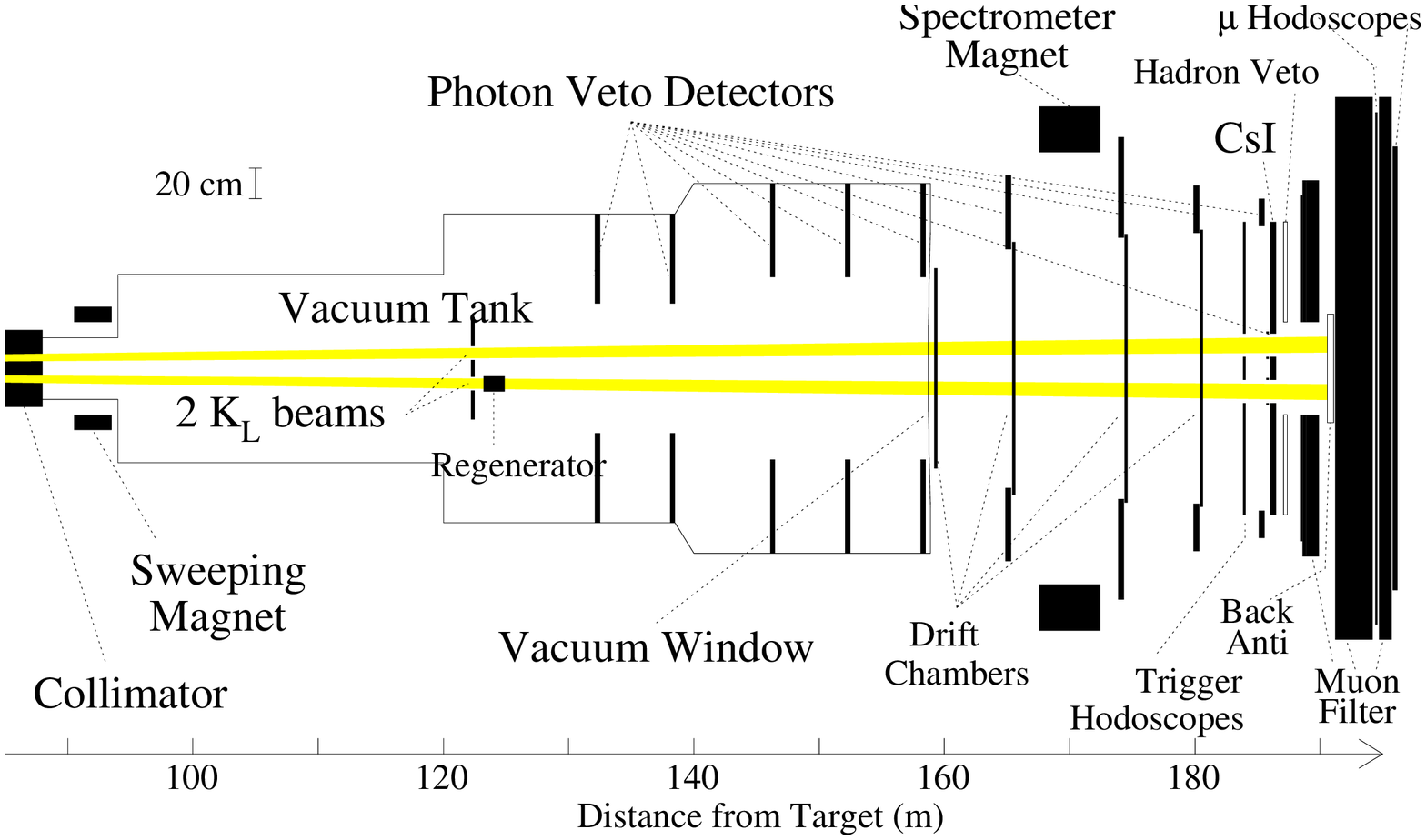,height=2.0in,width=3.5in}
\caption{The KTeV detector configured for measuring
Re($\epsilon'/\epsilon$). \label{detector}}
\end{figure}

The KTeV beamline and detector at Fermilab is described 
elsewhere\cite{pss}.  For this analysis, the detector is 
configured for the measurement of Re($\epsilon'/\epsilon$).
As shown in Fig. \ref{detector}, two approximately parallel
neutral $K_L$ beams enter a long 
vacuum tank, which defines the fiducial volume for accepted decays.  One 
of the beams strikes an active absorber (regenerator), which serves to tag 
the coherent regeneration of $K_S$.  The regenerator moves 
to the other beam in between Tevatron spill cycles.  
Following the vacuum tank, there are 4 planar drift chambers 
and an analysis
magnet that imparts a 411 MeV/c horizontal transverse 
kick to the charged particles.  
A high precision 3100-element pure Cesium Iodide calorimeter (CsI) 
is used primarily to measure the energy of $e^{\pm}$ and photons.  
Photon veto
detectors surrounding the vacuum tank, drift chambers, and CsI
serve to reject events with particles escaping the calorimeter.  

Only Ke3 decays in the beam opposite the regenerator
were used in 
this analysis.  They were triggered by the presence of hits in the
drift chambers and scintillators placed immediately upstream of the CsI, and
by the lack of activity in the regenerator or photon vetoes. 
To keep the event rate managable, the trigger rejected beam muons and
K$\mu3$ events by requiring no activity in the scintillation counters (muon
veto) placed behind 4 meters of Fe absorbers downstream of the CsI.

%\section{Analysis}

%\subsection{Ke3 Selection}

Selecting Ke3 events is relatively straightforward, as they are the most
copious $K_L$ decay mode (BR(Ke3) $\approx$ 0.39).  
We identify two-track vertices in the region between 90 and 160 meters from 
the target with 
kinematics inconsistent with $K_L \to \pi^+\pi^-\pi^0$ (K$\pi$3) and 
$\Lambda(\bar{\Lambda}) \to p\pi^-(\bar{p}\pi^+)$.  
The tracks are required to extrapolate to CsI energy deposits while 
maintaining sufficient clearance
from the inner and outer edges.  A track is 
identified as an electron if its momentum exceeds 5 GeV/c and its 
CsI energy deposit divided by its momentum
(E/P) exceeded 0.925. Pions are identified as having momentum exceeding
8 GeV/c and E/P less than 0.925.  The tighter pion momentum
requirement rejects K$\mu3$ background, since 8 GeV/c is the 
threshold for 
minimum-ionizing particles to penetrate through to the muon veto counters. 

To minimize background from target $K_S$ coherently interfering with
the $K_L$, the
event proper time ($\tau$) had to exceed 10.5 $K_S$ lifetimes ($\tau_S$). 
The $K_L$ momentum ($P_K$) reconstruction has ambiguities inherent 
to any 3-body decay with one unobserved particle. 
The longitudinal component of the neutrino momentum in the kaon rest
frame can be either parallel or anti-parallel to the kaon flight direction
in the lab frame.  This introduces two possible solutions for $P_K$. 
We use the low $P_K$
solution to calculate $\tau$.  Based on Monte Carlo (MC) simulations, this
choice is at least 70\% correct for $\tau < 22.5~\tau_S$.  
A small correction, derived from MC, was made for errors in the $P_K$ solution
and the residual target $K_S$ interference.

%\subsection{Acceptance and Efficiency Cancellation}

The measurement of $\delta_L$ requires a careful control of systematics.  
The most important of these is to ensure an equal
acceptance and efficiency for oppositely charged particles.  We 
combine data sets of opposite analysis magnet polarities, which 
was reversed about once per day, to ensure that the detector is exposed
to oppositely charged particles in the same manner.
We also used data collected during rare decay running (E799), which 
included a transition radiation detector 
(TRD)\cite{graham,solomey} used for the study of
electron and pion identification.

\begin{figure}[thb]
\vskip -1.0cm 
\hskip -0.5cm
\psfig{figure=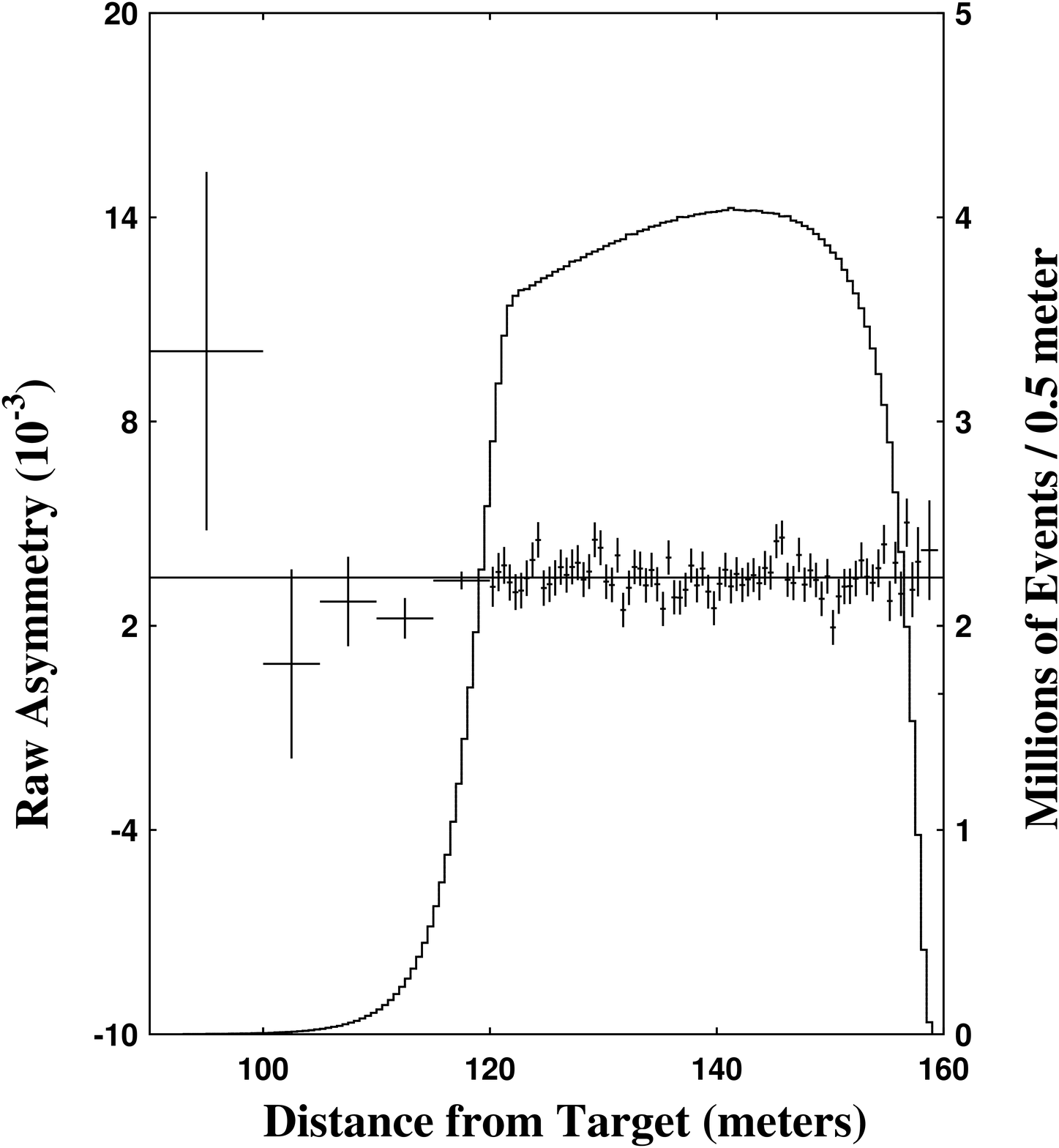,height=3.0in,width=3.5in}
\vskip -0.75cm
\caption{Raw $\delta_L$ and number of events 
versus vertex distance from target. A fit to a constant yields 
3417 $\pm$ 58 ppm with a $\chi^2$ of 81.53 for 81 degrees of 
freedom.\label{vtxz}}
\end{figure}

There are eight possible configurations for Ke3 decays: $e^+\pi^-$ or 
$e^-\pi^+$, east ($E$) or west ($W$) $K_L$ beams, 
and positive ($+$) or negative ($-$)
magnet 
polarities.  The number of Ke3 events in each configuration ($N_i$)
depends on the branching ratio ($BR$), the
acceptance and efficiency ($A$), and the 
flux $N(K_L)$.  We define $R$ as:
\begin{eqnarray}
\label{8fold}
R^4 & = &
      \frac{BR(e^+\pi^-) A(e^+\pi^-, E, +) N(K_L,E,+)}
           {BR(e^-\pi^+) A(e^-\pi^+, E, -) N(K_L,E,-)}  ~~~~~~~\\
& {\times} & \frac{BR(e^+\pi^-) A(e^+\pi^-, E, -) N(K_L,E,-)}
               {BR(e^-\pi^+) A(e^-\pi^+, E, +) N(K_L,E,+)} \nonumber \\
& {\times} & \frac{BR(e^+\pi^-) A(e^+\pi^-, W, +) N(K_L,W,+)}
               {BR(e^-\pi^+) A(e^-\pi^+, W, -) N(K_L,W,-)} \nonumber \\
& {\times} &  \frac{BR(e^+\pi^-) A(e^+\pi^-, W, -) N(K_L,W,-)}
               {BR(e^-\pi^+) A(e^-\pi^+, W, +) N(K_L,W,+)}, \nonumber 
\end{eqnarray}
where the four numerators and denominators represent 
$N_i,~i=1,8$.  Since the fluxes cancel and 
$A(e^\pm\pi^\mp)$ = $A(e^\mp\pi^\pm)$ under magnetic field
reversal,  Eq. \ref{8fold} reduces to:
\begin{eqnarray}
\label{8-fold}
{\rm raw~}\delta_L  =  \frac{R-1}{R+1} ~~~~~~
\sigma ({\rm raw}~\delta_L) =  
\frac{1}{8} \cdot \sqrt{ \Sigma_{i=1}^8 \frac{1}{N_{i}}}.
\end{eqnarray}
Equation 
\ref{8-fold} defines the ``raw'' $\delta_L$, and does not take into
account several effects discussed below. 
Our data yields raw $\delta_L~=~(3417~\pm~58)\cdot10^{-6}$ (ppm). 
Figure \ref{vtxz} shows its dependence on the 
decay vertex distance from the target.  

In principle, the same flux and acceptance/efficiency cancellation occurs 
in the pairing of the first and second (third and fourth) factors of 
Eq. \ref{8fold}, such 
that $\delta_L$ can be measured within the east and west beams.
Eq. \ref{8-fold} is the 
geometric mean of the two $\delta_L$ measurements.

\begin{figure}[thb]
\vskip -0.7cm
\hskip 0.75in
\psfig{figure=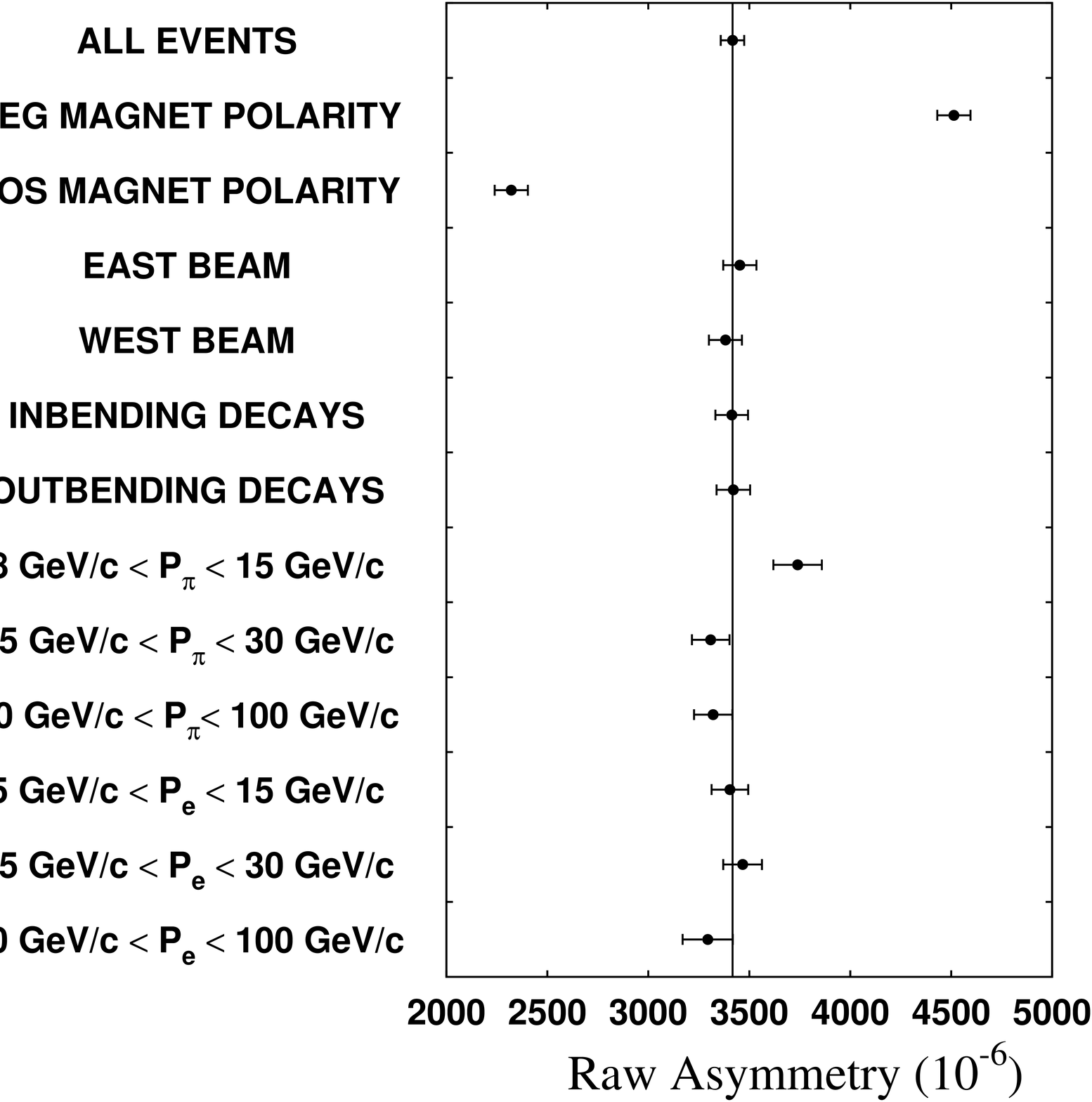,height=3.0in,width=3.1in}
\vskip -0.7cm  
\caption{Raw $\delta_L$ for all events and various 
data subsets. The line indicates the raw value of 
3417 ppm.\label{sys_check}}
\end{figure}

Figure \ref{sys_check} shows the raw asymmetry in various data subsets.
In particular, we see the importance of combining data from opposite
polarity runs. 
% To appreciate the importance of combining data from opposite polarity runs,
% Fig. \ref{sys_check} shows 
% the effect of using only one magnet setting.
The difference between measurements made with only one magnet
setting, $\delta_L(+) - \delta_L(-)$, is significant and on 
the order of $\delta_L$ itself.  This 
difference is due to very small offsets of 
the inner apertures and is well-reproduced by our MC.  
The MC is not used to correct this large effect since
Eq. \ref{8fold} naturally accounts for such geometrical asymmetries.
On the other hand, the values for $\delta_L$ agree between 
decays in the east and west beams.  Other data divisions yield good 
agreement except for the sample with 
$8~{\rm GeV/c} < P_{\pi} < 15~{\rm GeV/c}$ (discussed below).

% Two corrections to raw $\delta_L$, related to the magnetic field, 
% were considered:
% (1) that the field was not reversed exactly, and (2) that the 
% detector itself behaved differently between polarity settings.
% The field magnitudes $|B_x|$ and $|B_y|$ differed
% by $\approx$ 0.4 MeV/c equivalent between opposite polarities.
% The $B_y$ misreversal is 
% consistent with contribution from the earth's field.
% The $B_x$ misreversal was traced to the magnet behaviour 
% itself and confirmed by zip-track measurements.  The magnet misreversal
% correction was determined from MC to be $-3.1 \pm 1.6$ ppm.
% The detector itself may behave differently between polarity 
% settings; this effect is expected to be quite small 
% since the stray fields are less than 10 gauss at the trigger hodoscope and
% CsI pmts.  In addition, we found no significant dependence of the electron 
% and pion efficiency on magnet polarity.
 
%\subsection{Material Systematics}

We account for the different behavior of particles and antiparticles in
matter.  For each cut that introduces a bias, we 
measure $f^+$ ($f^-$), the inefficiency 
in the $e^+\pi^-$ ($e^-\pi^+$) configuration.
The correction to raw $\delta_L$ for 
the cut, summarized in table \ref{summary}, is simply
$(f^+ - f^-)/2$. 

Many biases due to $\pi^{\pm}$ interaction differences were considered, 
and data were used to measure most of these corrections.  
These include the $\pi^{\pm}$ energy deposition in the CsI, 
the loss of pions due to interactions in the trigger scintillators,
and pion punchthrough past the 
Fe absorbers depositing energy in the muon veto.  
Events with E/P for pions exceeding 0.925 would 
be removed since the analysis requires
exactly one identified electron.   This effect, studied using pions in
K$\pi$3 events, removes
$\approx 0.6\%$  of the pions, with the probability being slightly
larger for $\pi^+$ than $\pi^-$.  The asymmetry between $\pi^+$ and $\pi^-$
has a strong momentum 
dependence and explains the trend seen for the 
$8~{\rm GeV/c} < P_{\pi} < 15~{\rm GeV/c}$ data sample 
(see Fig. \ref{sys_check}). The correction to $\delta_L$ is 
$-156 \pm 10$ ppm.  

A correction was estimated for possible biases due to the trigger muon 
veto requirement, which removed approximately 3\% of pions due to 
decay-in-flight and punchthrough.  Good events can also be lost if 
accompanied by accidental muons.
% While decay-in-flight and accidental losses should introduce
% no bias,  the neutron excess in the Fe absorbers would bias pion 
% punchthrough.  
Only pion punchthrough would cause a bias due to the neutron
excess in the Fe absorbers.  The estimate used a
Ke3 sample collected in E799, where the TRD gave additional electron
purity. The sample is 
triggered by {\it requiring} muon counter hits, as these events 
are the ones that would be removed by the analysis trigger.
This sample had
a charge asymmetry of $(4.4 \pm 1.3)\cdot 10^{-3}$, consistent 
with no bias.   Nevertheless, we use a correction
of $34 \pm 40$ ppm, where the error is given by the sample statistics.

Pion interactions in the trigger scintillator can
confuse the reconstruction so that the track fails to match to a CsI
cluster. This effect, studied using Ke3 events with a relaxed track-cluster 
matching requirement, removes $\approx$ 0.42\% of all pions and 
was slightly more probable for the $\pi^-$ than $\pi^+$, leading to a 
correction of $54 \pm 10$ ppm.  

Considering $e^{\pm}$ interaction differences, one has primarily $e^+$ 
annihilation and $\delta$-ray production. An 
$e^+$ annihilation occuring near the upstream surface of the trigger
scintillator would fail the trigger requirement
due to the reduced energy deposition.
$\delta$-rays more than 10 MeV and emitted from the vacuum window or 
tracking system can cause losses since the 
momentum transfer would drastically
change the event kinematics.  These losses cause a bias due to the small
difference between the Bhabha ($e^+e^-$) and Moller ($e^-e^-$) scattering 
cross section. 
These corrections were derived by a Geant simulation of the detector material.
They are of order $\pm$10 ppm, mitigated by the high momentum of the 
tracks and the minimal material (2\% $X_0$) upstream of the CsI.

Finally, a Ke3 sample was used to study potential biases of the E/P cut on 
electrons.  This sample was selected by a relaxed E/P cut and by the tagging
of a minimum-ionizing $\pi$ in the CsI.  The E/P removed removed approximately 
0.233\% of $e^+$ and $e^-$.  No bias is seen, and we assigned a correction of 
$-19 \pm 18$ ppm.  This result 
was confirmed with a similar Ke3 sample collected in E799,
where the TRD gave additional electron purity.

% A small correction was attributed to pions absorbed in the vacuum window, 
% drift chambers, and helium bags.  
% This was estimated assuming isospin conservation in strong interactions,
% and so the $\pi^+$ and $\pi^-$ absorption differences depended on the 
% excess protons, which was predominantly in the form of hydrogen.
% We used the {\it total} cross section $\sigma_T(\pi^+ p)$ and 
% $\sigma_T(\pi^- p)$ listed in \cite{pdg}.  
% Because we assume that these hadronic interactions
% removed the events from analysis, which was not always the case, 
% we attribute a 100\% uncertainty to this correction. 
%
% An estimate was made for bias due to $\pi^-$ charge exchange near the 
% upstream surface of the trigger scintillator.  In this process, the  
% event would fail the trigger due to the $\pi^-$ reduced energy deposit.
% Measurements of energy deposit were performed for 1 GeV $\pi^-$ in 
% plastic scintillator \cite{inagaki}.  Using these measurements and 
% \cite{flaminio} to extrapolate to our $\pi^-$ momentum range, we find a 
% correction of 2 $\pm$ 1 ppm.  

We applied a small correction to account for 
pions absorbed in the vacuum window, 
drift chambers, and helium bags.  
This was estimated assuming isospin conservation in strong interactions,
and so the $\pi^+$ and $\pi^-$ absorption differences depended on the 
excess protons, which was predominantly in the form of hydrogen.
We accounted for a small bias due to $\pi^-$ charge exchange near the 
upstream surface of the trigger scintillator (analogous to the case of 
$e^+$ trigger loss). For this, we used the measurements of \cite{inagaki} and 
\cite{flaminio} and extrapolated them to our $\pi^-$ momentum range.

\begin{figure}[thb]
% \vskip -1.0cm
% \psfig{figure=world-summary-prl.eps,height=2.4in,width=3.5in}
\psfig{figure=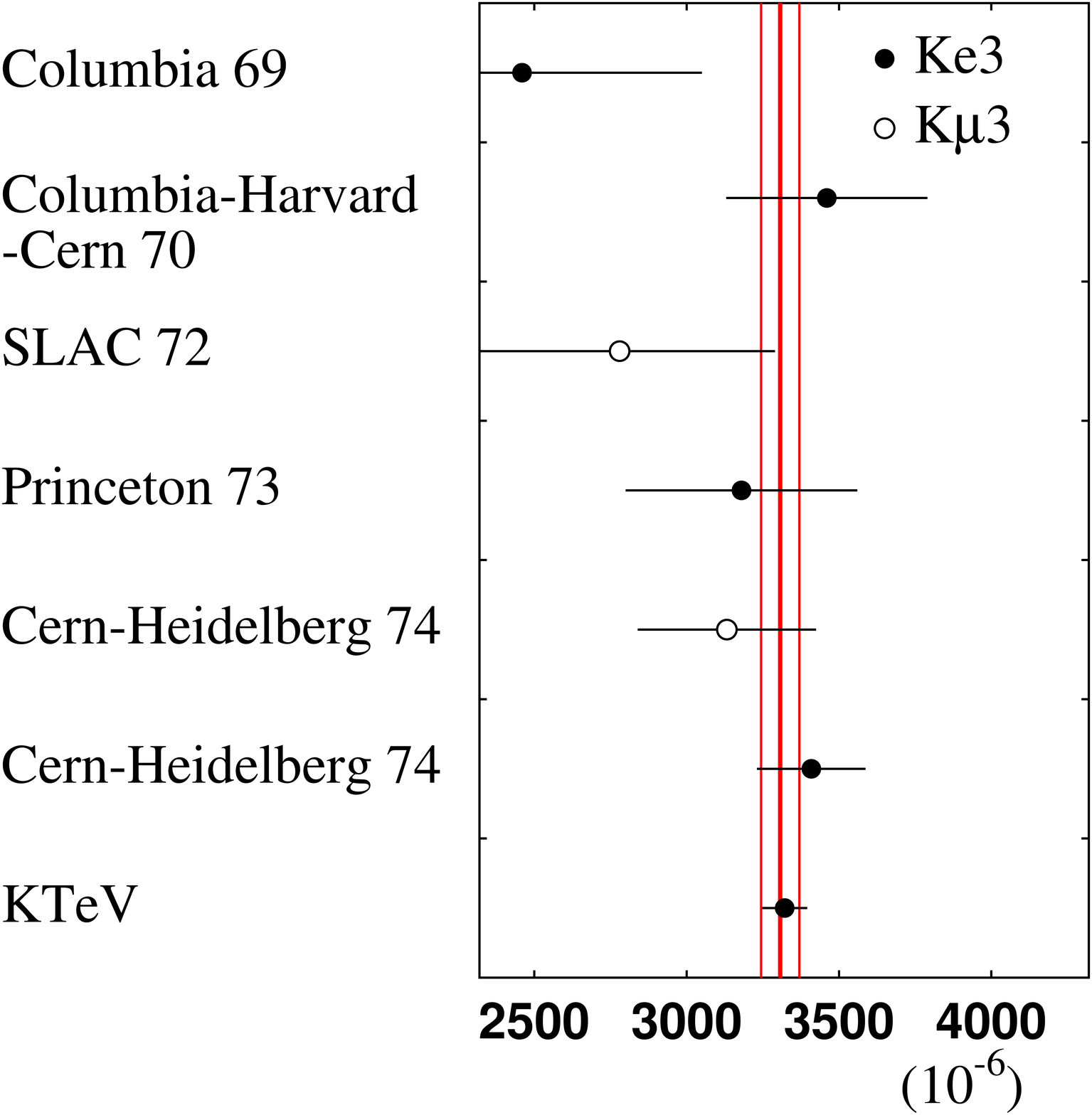,height=2.8in,width=3.5in}
\vskip -0.75cm
\caption{Compilation of $\delta_L$ measurements including this 
result. Lines indicate the average and its uncertainty.\label{combined}}
\end{figure}

Other small corrections, derived from MC, accounted for the inexact reversal 
of the analysis magnet polarity and the residual coherent 
$K_S$ from the target, absorber, and $K_L$ scatters with the
final collimator and regenerator.  
%
% The effect of non-Ke3 backgrounds is found to be almost negligible.
%
\begin{table}[htpb]
\caption{Corrections in ppm for this analysis. \label{summary}}
\begin{center}
\begin{tabular}{lc} 
\hline
$\pi^+\pi^-$ difference in CsI    & $ -156 \pm 10$  \\
$\pi^{\pm}$ interaction in trigger scintillator &  $+54 \pm 10$ \\
$\pi$ decay and punchthrough    &    $+34 \pm 40$  \\
$e^{+}e^{-}$ difference in CsI  &  $ -19  \pm 18 $  \\
target/absorber $K_S - K_L$ interference & $ -12 \pm 1 $ \\
$e^+$ annihilation in spectrometer & $ +11 \pm 1$  \\
Delta ray production   &  $ -8.5 \pm 4.3 $ \\
$\pi$ absorption in spectrometer   & $+5.0 \pm 3.2$ \\
inexact analysis magnet polarity reversal   &  $ -3.1 \pm 1.6 $ \\
final collimator and regenerator scatters &  $ -1.2 \pm 2.3 $  \\
$\pmz$ background               &   $ +0.5 \pm 0.7 $   \\ 
$K{\mu}3$ background            &    0 $\pm$ 0 \\
$\Lambda_{p\pi}$ and $\Lambda_{\beta}$  background & 0 $\pm$ 0 \\
% Regenerator crossovers  & 0 $\pm$ 0 \\
\hline 
Total correction    &  $ -95.3 \pm 46.5 $
\end{tabular}
\end{center}
\end{table}
%\section{Result and Conclusion}
Table \ref{summary} shows the summary of all systematic corrections.  The
uncertainties are uncorrelated since they are mainly statistical errors of the
Ke3 and K$\pi$3 control samples.
Combining the corrections with raw $\delta_L$, we find:
\begin{eqnarray}
\delta_L & = & 3322 \pm 58~(stat) \pm 47~(syst)~~{\rm ppm} \nonumber \\
         & = & 3322 \pm 74~(combined)~~{\rm ppm.} \nonumber 
\end{eqnarray}
This result is in excellent agreement with previous measurements 
and 2.4 times 
more precise than the current best result (see Fig. \ref{combined}).
A combination of all results including ours yields:
\begin{eqnarray}
\delta_L  = 3307 \pm 63~~{\rm ppm} ~~~~~ 
\chi^2 & = & 4.2/6~~{\rm d.o.f..} \nonumber 
\end{eqnarray}
% The K$\pi$2 amplitude ratios ($\eta_{+-}, \eta_{00}$) 
% and $\delta_L$ can be used to place
% limits on CPT violation.  Using $\eta_{+-}$ and $\eta_{00}$ values 
% from \cite{pdg} and the combined $\delta_L$:
Substituting $\eta_{+-}$ and $\eta_{00}$ values 
from \cite{pdg} and the combined $\delta_L$ into Eq. \ref{cpt}, we find:
\begin{eqnarray}
% \label{cpt}
Re~(Y + \frac{x - \bar{x}}{2} + a) % & = & Re(\frac{2}{3}\eta_{+-} + \frac{1}{3}\eta_{00}) - \frac{\delta_L}{2} \nonumber \\
 &  = & (1650 \pm 16) - (1653 \pm 32)\nonumber \\
 &  = & -3 \pm 35 ~ {\rm ppm.} \nonumber 
\end{eqnarray} 
% where $Re~a$ parameterizes CPT violation in K$\pi$2
% $K^0(\overline{K^0}) \rightarrow 2\pi$ 
% decays\cite{barmin}.
The result is consistent with no CPT violation and is limited by the charge
asymmetry uncertainty.  It limits 
$|Re~(Y + (x - \bar{x})/2 + a)| < 61~{\rm ppm}$ at the 90\% C.L..
Barring fortuitous cancellations, this is thus far the most stringent limit on 
$Y,~x,~{\rm and}~a$.

\begin{acknowledgments}
  We gratefully acknowledge the support and effort of the Fermilab
  staff and the technical staffs of the participating institutions for
  their vital contributions.  This work was supported in part by the
  DOE, the NSF and the
  Ministry of Education and Science of Japan.  A.R.B.,
  E.B. and S.V.S.  acknowledge support from the NYI program of the
  NSF; A.R.B. and E.B. from the Alfred P. Sloan Foundation; E.B. from
  the OJI program of the DOE; K.H., T.N. and M.S. from the Japan
  Society for the Promotion of Science; and R.F.B from the
  Funda\c{c}\~{a}o de Amparo \`{a} Pesquisa do Estado de S\~{a}o
  Paulo.  P.S.S. acknowledges receipt of a Grainger Fellowship.\\
$^{\dagger}$ To whom correspondence should be addressed: hogann@fnal.gov. \\
$^{*}$ Perm. address C.P.P. Marseille/C.N.R.S., France \\
$^{**}$Perm. address Univ. of S\~{a}o Paulo, S\~{a}o Paulo, Brazil
\end{acknowledgments}

\end{document}